\documentclass[aps,preprint,showpacs,groupedaddress,floatfix]{revtex4}%
\usepackage{graphicx}
\usepackage{dcolumn}
\usepackage{bm}
\usepackage{amsmath}
\usepackage{amsfonts}
\usepackage{amssymb}%
\providecommand{\U}[1]{\protect\rule{.1in}{.1in}}

\newcommand{\be}{\begin{equation}}
\newcommand{\ee}{\end{equation}}
\newcommand{\bea}{\begin{eqnarray}}
\newcommand{\eea}{\end{eqnarray}}

\begin{document}
\title{Relativistic quasiparticle time blocking approximation.
II. Pygmy dipole resonance in neutron-rich nuclei}
\author{E. Litvinova}
\affiliation{Gesellschaft f\"{u}r Schwerionenforschung mbH, 64291
Darmstadt, Germany} \affiliation{Frankfurt Institute for Advanced
Studies, Universit\"at Frankfurt, 60438 Frankfurt am Main,
Germany} \affiliation{Institute of Physics and Power Engineering,
249033 Obninsk, Russia}
\author{P. Ring}
\affiliation{Physik-Department der Technischen Universit\"at M\"unchen, D-85748 Garching, Germany}
\author{V. Tselyaev}
\affiliation{Nuclear Physics Department, V. A. Fock Institute of Physics, St. Petersburg
State University, 198504, St. Petersburg, Russia}
\author{K. Langanke}
\affiliation{Gesellschaft f\"{u}r Schwerionenforschung mbH, 64291
Darmstadt, Germany} \affiliation{Institut f\"ur Kernphysik,
Technische Universit\"at Darmstadt, 64291 Darmstadt, Germany}
\date{\today}

\begin{abstract}
Theoretical studies of low-lying dipole strength in even-even
spherical nuclei within the relativistic quasiparticle time blocking
approximation (RQTBA) are presented. The RQTBA developed recently as
an extension of the self-consistent relativistic quasiparticle random
phase approximation (RQRPA) enables one to investigate effects of
coupling of two-quasiparticle excitations to collective vibrations
within a fully consistent calculation scheme based on covariant
energy density functional theory. Dipole spectra of even-even
$^{130}$Sn -- $^{140}$Sn and $^{68}$Ni -- $^{78}$Ni isotopes
calculated within both RQRPA and RQTBA show two well separated
collective structures: the higher-lying giant dipole resonance (GDR)
and the lower-lying pygmy dipole resonance (PDR) which can be
identified by a different behavior of the transition densities of
states in these regions.
\end{abstract}

\maketitle

\section{Introduction}

Since the last decade experimental and theoretical investigations of
the pygmy dipole resonance (PDR) remain a challenge of nuclear
structure physics. Due to recent developments of experimental
techniques, in particular, the high resolution nuclear resonance
fluorescence, valuable information has been obtained about low-lying
dipole spectra of stable nuclei~\cite{GBB.98,Rye.02,ZVB.02}.
Pioneering Coulomb dissociation experiments~\cite{Adr.05,KPA.07}
allowed to extend investigations of the low-lying dipole strength to
nuclei with large neutron excess.


Properties of the pygmy dipole resonance obtained in the various
experimental and theoretical studies can provide unique information
about nuclear structure. It has been pointed out that the pygmy
dipole strength is closely related to static nuclear properties, for
instance, to the neutron skin thickness and the nuclear symmetry
energy~\cite{TL.08,KPA.07}. Besides the pure nuclear structure
interest, an existence of a PDR just above the neutron threshold can
have significant astrophysical importance as it may increase the
neutron capture cross sections essential for r-process
nucleosynthesis simulations (see, for instance, \cite{GKS.04} and
references therein).

So far the best description of the experimentally observed low-lying
dipole strength in stable medium and heavy nuclei is achieved in the
calculations within the quasiparticle-phonon model
(QPM)~\cite{Sol.92}. The first $1^-$ state, the position, the total
strength and the fine structure of the PDR are described very well in
the wide model space including up to three-phonon
configurations~\cite{GBB.98,Rye.02}. However, because of lack of
self-consistency and the presence of adjustable parameters, the
application of this approach to nuclei with large neutron excess
remains questionable. Another successful tool to investigate the
low-energy dipole response are approaches based on the covariant
energy density functional, first of all, the fully self-consistent
relativistic quasiparticle random phase approximation
(RQRPA)~\cite{PRN.03}, see also Ref.~\cite{PVK.07} for a recent
review. The RQRPA supplemented with the coupling to low-lying
vibrations within the relativistic quasiparticle time blocking
approximation (RQTBA)~\cite{LRT.08} in a fully consistent way enables
one to reproduce the fragmentation of the giant dipole resonance as
well as of the PDR and to describe the dipole strength of the
low-energy part of the spectrum. The coupling of the single-particle
motion to vibrations is known as an important mechanism for the
damping of particle-hole excitations. Taking into account this
mechanism is necessary to describe the widths of giant
resonances~\cite{BBBD.79,SBC.04,KST.04,Tse.07,LT.07}. Since the PDR
is substantially a nuclear surface phenomenon, its coupling to
surface vibrations may be significant.

The present paper is considered to be a contribution to the
understanding of properties of the pygmy dipole resonance in
neutron-rich nuclei. We investigate the dipole response of even-even
spherical nuclei with large neutron excess within the RQTBA developed
recently in Ref.~\cite{LRT.08} paying special attention to the lowest
part of the dipole spectrum. Relying on the previous
calculations~\cite{LRT.07,LRT.08}, which demonstrate very good
agreement with data for the experimentally known nuclei, we retain
the same calculation scheme and do not introduce any additional
parameters or approximations.

\section{RQTBA calculations: agreement with data and predictive power}

In Ref.~\cite{LRT.08} the relativistic quasiparticle time blocking
approximation has been developed and applied for calculations of the
dipole response of medium-mass open-shell nuclei. The physical
content of this approach is the quasiparticle-vibration coupling
model based on covariant energy density functional theory (CEDFT) and
the relativistic QRPA (RQRPA)~\cite{PRN.03}. To describe excitations
in nuclei with even particle number the Bethe-Salpeter equation (BSE)
containing both static and energy-dependent residual interactions has
been formulated for the nuclear response function $R(\omega)$ in the
doubled quasiparticle space:
\begin{equation}
R_{k_{1}k_{4},k_{2}k_{3}}^{\eta\eta^{\prime}}(\omega)=\tilde{R}_{k_{1}k_{2}%
}^{(0)\eta}(\omega)\delta_{k_{1}k_{3}}\delta_{k_{2}k_{4}}\delta_{\eta
\eta^{\prime}}+\tilde{R}_{k_{1}k_{2}}^{(0)\eta}(\omega)\sum\limits_{k_{5}%
k_{6}}\sum\limits_{\eta^{\prime\prime}}{\bar{W}}_{k_{1}k_{6}%
,k_{2}k_{5}}^{\eta\eta^{\prime\prime}}(\omega)R_{k_{5}k_{4},k_{6}k_{3}}%
^{\eta^{\prime\prime}\eta^{\prime}}(\omega),
\label{respdir}%
\end{equation}
where the indices $k_i$ run over the single-particle quantum numbers
including states in the Dirac sea and indices $\eta, \eta^{\prime},
\eta^{\prime\prime}$ numerate forward (+) and backward (-) components
in the doubled quasiparticle space. The quantity
$\tilde{R}^{(0)}(\omega)$ describes the free propagation of two
quasiparticles in the mean field between their interaction with the
amplitude
\begin{equation}
{\bar{W}}_{k_{1}k_{4},k_{2}k_{3}}^{\eta\eta^{\prime}}(\omega)=\tilde{V}%
_{k_{1}k_{4},k_{2}k_{3}}^{\eta\eta^{\prime}}+\Bigl(\Phi_{k_{1}k_{4},k_{2}%
k_{3}}^{\eta}(\omega)-\Phi_{k_{1}k_{4},k_{2}k_{3}}^{\eta}(0)\Bigr)\delta
_{\eta\eta^{\prime}}.
\label{W-omega}%
\end{equation}
The static part of the single-quasiparticle self-energy, determined
by the CEDFT with the parameter set NL3 \cite{NL3}, is based on a
one-meson exchange interaction with a non-linear self-coupling
between the mesons. Pairing correlations, considered to be a
non-relativistic effect and treated in terms of Bogoliubov's
quasiparticles within the BCS approximation, are introduced into the
relativistic energy functional as an independent phenomenologically
parameterized term.  The energy-dependent part of the
single-quasiparticle self-energy is modelled by the
quasiparticle-phonons coupling (QPC). The energy-dependent residual
interaction $\Phi(\omega)$ is derived from the energy-dependent
self-energy by the consistency condition and calculated within the
quasiparticle time blocking approximation. This approximation means
that, due to the time projection in the integral part of the BSE, the
two-body propagation through states which have a more complicated
structure than $2q\otimes phonon$ has been blocked~\cite{Tse.07}. In
order to avoid double counting of the QPC effects a proper
subtraction of the static QPC contribution from the residual
interaction has been performed in Eq.~(\ref{W-omega}), since the
parameters of the CEDFT have been adjusted to experiment and include
already essential ground state correlations. In order to take the QPC
into account in a consistent way, we have first calculated the
amplitudes of this coupling within the self-consistent RQRPA with the
static residual interaction $\tilde V$. Then, the BSE (\ref{respdir})
for the response function has been solved in both Dirac-Hartree-BCS
and momentum-channel representations.

To describe the observed spectrum of an excited nucleus in a weak
external field $P$ as, for instance, an electromagnetic field, the
microscopic strength function $S(E)$ is computed:
\begin{equation}
S(E) = -\frac{1}{2\pi}\lim\limits_{\Delta\rightarrow+0}Im\
\sum\limits_{k_{1}k_{2}k_{3}k_{4}}\sum\limits_{\eta\eta^{\prime}}
P_{k_{1}k_{2}}^{\eta\ast}R_{k_{1}k_{4},k_{2}k_{3}}^{\eta\eta^{\prime}}
(E + i\Delta)P_{k_{3}k_{4}}^{\eta^{\prime}}.
\label{strf}%
\end{equation}
In the calculations a finite imaginary part $\Delta$ of the energy
variable is introduced for convenience in order to obtain a
smoothed envelope of the spectrum. This parameter has the meaning
of an additional artificial width for each excitation. This width
emulates effectively contributions from configurations which are
not taken into account explicitly in our approach. The dipole
photoabsorption cross section $\sigma_{E1}(E)$ is expressed via
the strength function $S_{E1}(E)$ according to the well known
formula:
\be \sigma_{E1}(E) = \frac{16 \pi^3 e^2}{9{\hbar}c} E S_{E1}(E)\,.
\label{sgth}
\ee

In Ref.~\cite{LRT.08} numerical results for dipole spectra in tin
isotopes and N=50 isotones have been presented, which were obtained
within the approach described above and compared with RQRPA
calculations and the available experimental data. The mean energies
(ME) and the energy-weighted sum rule (EWSR) of the giant dipole
resonance (GDR) in all nuclei under investigation agree very well
with the experimental data, and this is provided by the
self-consistent RQRPA. The coupling to phonons is responsible for the
fragmentation of the resonance. In our fully consistent approach it
does almost not affect the nice agreement achieved in RQRPA for the
ME and the EWSR in the GDR region.

Notice, that in stable nuclei not only the integral characteristics
like the mean energy, the energy-weighted sum rule and the width, but
also the general shape of the GDR are obtained in the RQTBA rather
close to those observed in experiments demonstrating the new quality
of description as compared to previous microscopic studies. Thus,
from Ref.~\cite{LRT.08}, where an excellent agreement has been
achieved with the data for the GDR in $^{116}$Sn, $^{120}$Sn,
$^{88}$Sr, $^{90}$Zr, we conclude that the main mechanisms of the GDR
formation in medium mass and heavy nuclei are taken into account
correctly and consistently in the RQTBA. Moreover, it turned out to
be possible to obtain a very good description of the low-lying part
of the dipole strength below the nucleon separation energy in the
same calculation scheme. Strictly speaking, a correct comparison of
the integrated pygmy dipole strength with experimental data in this
region is still an open problem since the existing data, which result
in most cases from the ($\gamma,\gamma^{\prime}$) measurements, are
usually restricted from the top by the nucleon separation energies
because of loss of experimental sensitivity. In theoretical
calculations beyond (Q)RPA the pygmy mode is in most cases strongly
coupled to the surface vibrations which causes the spreading of the
low-lying strength with the isoscalar underlying structure in a broad
energy region, for instance, 3-10 MeV for tin isotopes with medium
neutron excess. So, at present the comparison with data is only
possible for the lowest part of the pygmy strength measured below 8
MeV. For QRPA calculations the situation is even more difficult: in
particular, in self-consistent RQRPA calculations~\cite{PRN.03} only
one or very few pronounced peaks, corresponding to the soft dipole
mode, occur and accumulate the whole strength with the isoscalar
underlying structure, so that they often do not overlap at all with
the experimentally investigated region. However, the situation is
different for the RQTBA approach. Within this model~\cite{LRT.08} the
integrated dipole strength below 8 MeV is in very good agreement with
the data which are available for $^{88}$Sr~\cite{SRB.07} and
$^{116}$Sn~\cite{GBB.98}. Note, that the coupling to phonons is the
essential mechanism which brings the strength to this region while
the RQRPA yields no dipole strength at all in this energy range.

\begin{figure}[ptb]
\begin{center}
\includegraphics*[scale=1.0]{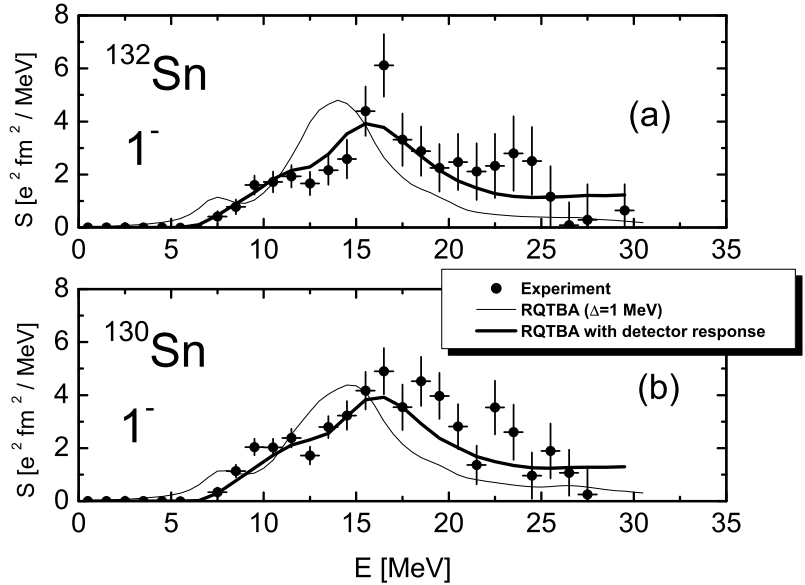}
\end{center}
\caption{Spectra of dipole excitations in $^{130}$Sn, $^{132}$Sn.
The experimental data \cite{Adr.05} are shown by the black circles with the error bars,
the thin curves represent our RQTBA calculations with 1 MeV smearing, and
the thick curves give the RQTBA strength distributions after the convolution
with the detector response \cite{KBA}.}
\label{f1}%
\end{figure}

Thus, the RQTBA has been successfully tested and demonstrated  as a
reliable approach for the description of excited states in stable
nuclei in the GDR as well as in the PDR regions. Since it contains no
adjustable parameters and uses only the universal relativistic mean
field (RMF) parameters, the approach should also give reliable
results for nuclei further away from the valley of stability. This
expectation has been indeed confirmed by a comparison with recent
pioneering experiments on Coulomb dissociation of
$^{130,132}$Sn~\cite{Adr.05}. The dipole strength calculated within
the RQTBA for the doubly magic $^{132}$Sn has been presented in
Ref.~\cite{LRT.07}, and its low-lying part has been analyzed in
Ref.~\cite{LRV.07}. Calculations for $^{130}$Sn have been done and
presented in Ref.~\cite{LRT.08}. However, a direct comparison of
these results to the data of Ref.~\cite{Adr.05} is not possible, but
rather requires a convolution of the calculated strength with the
detector response matrix~\cite{KBA}. For this purpose we have first
recomputed the dipole strength in these two nuclei with the smearing
parameter 1 MeV which corresponds to the experimental energy bin size
(note, that in Refs.~\cite{LRT.07,LRV.07,LRT.08} the calculations
were made with 200 keV smearing). The results are presented by the
thin solid curves in Fig.~\ref{f1}. The thick solid curves give the
results obtained after the convolution with the detector response.
The black circles with error bars are the experimental data. After
the convolution, the RQTBA dipole strength distributions reproduce
the data very well, although the major part of the RQTBA pygmy
strength resides below the neutron threshold, where the dipole
strength is not detectable in the experimental set-up of
Ref.~\cite{Adr.05}. Moreover, it turns out that, after the
convolution, the data can be similarly well reproduced by quite
distinct initial dipole strength distributions; e.g. by those
obtained within QRPA based on the Skyrme functional~\cite{TL.08}.
Thus, one may conclude that the data are only weakly sensitive to the
details of the original strength distribution and, for a better
understanding of the PDR phenomenon in these nuclei more precise data
are needed.

\section{Pygmy dipole resonance: general features}
\label{pygmygen}

Ref.~\cite{LRV.07} reports about detailed investigations of low-lying
dipole strengths in nuclei with closed shells within the relativistic
time blocking approximation (RTBA). It has been confirmed that the
PDR arises from a resonant oscillation of the neutron skin against
the isospin saturated proton-neutron core, with the corresponding
relativistic RPA state characterized by a coherent superposition of
many neutron particle-hole configurations. Moreover, this picture
remains essentially unchanged when the particle-vibration coupling is
included within the RTBA. The effect of two-phonon admixtures in
nuclei with closed shells results in a weak fragmentation and a small
shift of the PDR states to lower excitation energies, while the
one-phonon character and the underlying structure of the states are
not noticeably modified by the coupling to low-lying surface
vibrations.

However, calculations performed in Ref.~\cite{LRT.08} have shown that
the presence of pairing correlations at least in one of the nucleonic
subsystems may change this conclusion crucially. First of all, these
correlations lead to a diffuseness of the Fermi surface and, thus,
increase the number of possible configurations of $2q\otimes phonon$
type. But the main reason for the stronger coupling effect is the
considerable lowering of the energies and/or the increased transition
probabilities of the lowest 2$^+$ states due to pairing correlations
of the superfluid type. In spherical open-shell medium-mass nuclei
highly collective first 2$^+$ states appear at energies around 1 MeV
(and they are usually well reproduced in RQRPA) whereas in magic
nuclei and often in nuclei near shell closures they appear much
higher, at about 3-4 MeV and usually have a considerably reduced
strengths. This has important physical consequences resulting in
strong configuration mixing if low-lying vibrational states are
present and the fragmentation caused by this mixing is the stronger
the lower the excitation energies and the higher the transition
probabilities. This is confirmed by the calculations of the
Ref.~\cite{LRT.08}.

In the RQTBA calculations we can, in particular, analyze
admixtures to the pygmy mode which are associated with the
collective vibrational motion of the neutron or proton excess
against the isospin saturated core. This collective state, whose
collectivity is directly dependent on the neutron excess (i.e. the
$|$N-Z$|$ value), appears in the RQRPA as a pure mode with a
pronounced dominance of the neutron or proton component of the
transition densities at the nuclear surface, which, in turn, is
subject to vibrational motion. As a consequence, the pygmy mode is
not observed as a pure mode, but always couples to the surface
vibrations.

The results obtained in Ref.~\cite{LRT.08} for Z=50 and N=50 nuclei
with medium neutron excess also show that coupling to low-lying
vibrational states is the main mechanism which brings the dipole
strength to the low-energy region far below the neutron threshold,
whereas in pure RQRPA only one collective peak (or several peaks in
nuclei with larger neutron excess) around the threshold occurs. This
peak, which arises as a result of a pushing-down effect and the
coherence of the unperturbed two-quasiparticle states by the
isoscalar part of the effective meson-exchange interaction, then
spreads over a broad energy range due to phonon coupling. As a
result, one finds its fragments well below and above its original
position in close resemblance of the spreading mechanism of the GDR.

\section{Low-lying dipole strength in neutron-rich tin isotopes}
\label{sn} 
\begin{figure}[ptb]
\begin{center}
\includegraphics*[scale=1.2]{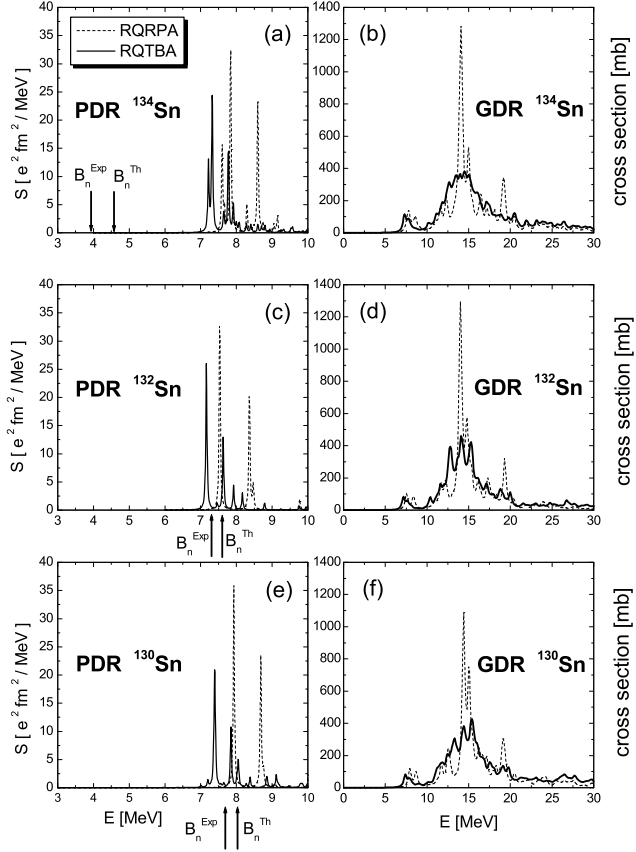}
\end{center}
\caption{The calculated dipole spectra for $^{130,132,134}$Sn isotopes. Right
panels: photoabsorption cross sections computed with the artificial width 200
keV. Left panels: the low-lying portions of the corresponding spectra in terms
of the strength function, calculated with 20 keV smearing. Calculations within
the RQRPA are shown by the dashed curves, and the RQTBA - by the solid curves.
Arrows denote the experimental and the theoretical neutron separation energies.}
\label{f2}%
\end{figure}

\begin{figure}[ptb]
\begin{center}
\includegraphics*[scale=1.2]{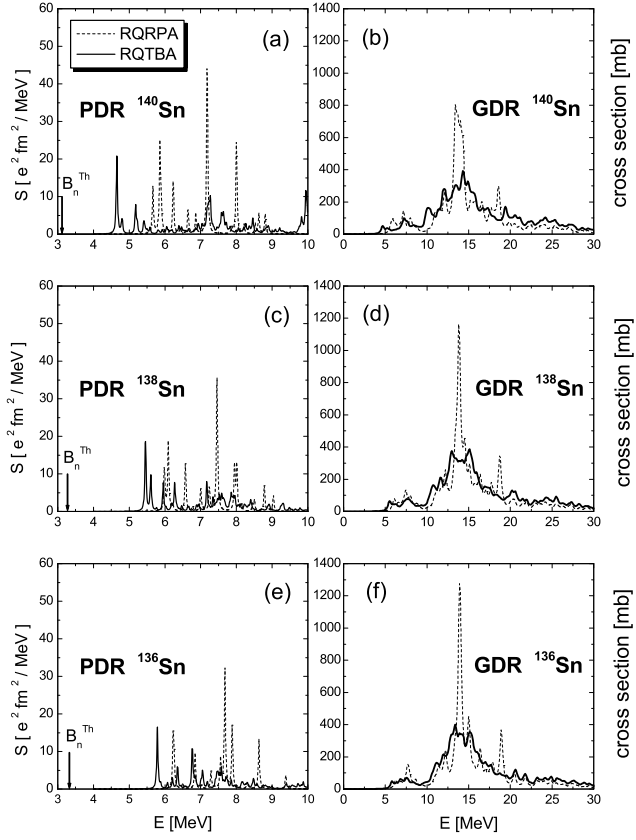}
\end{center}
\caption{Same as in Fig. \ref{f2}, but for $^{136,138,140}$Sn isotopes.}
\label{f3}%
\end{figure}

In Ref.~\cite{LRT.08} the dipole response in the tin
isotopic chain has been investigated up to $^{130}$Sn (with neutron
number N=80), while results for the doubly magic nucleus $^{132}$Sn
are given in Refs.~\cite{LRT.07,LRV.07}. Here we extend these studies
of the dipole response in even-even tin isotopes to calculation of
the very neutron-rich isotopes $^{134}$Sn, $^{136}$Sn, $^{138}$Sn,
$^{140}$Sn. The results are shown in Figs.~\ref{f2} and~\ref{f3},
including also results obtained previously for $^{130}$Sn and
$^{132}$Sn for comparison. The right panels of the figures display
the photo absorption cross sections, calculated with the usual
isovector dipole operator, using a 200 keV smearing. The left panels
enlarges the low-lying parts of the corresponding spectra, calculated
with a smaller imaginary part for the energy variable (20 keV), in
order to pronounce the fine structure of the spectra and to even
identify individual levels in this energy region. Calculations within
the RQRPA and RQTBA are shown by the dashed and solid curves,
respectively. The arrows denote the positions of the experimental and
the theoretical neutron thresholds, showing reasonably good
agreement.

Fig.~\ref{f2} represents the case of the doubly magic nucleus
$^{132}$Sn and its neighboring even-even nuclei $^{130}$Sn and
$^{134}$Sn. Comparing these results to those obtained in
Ref.~\cite{LRT.08} for $^{116}$Sn and $^{120}$Sn and with the more
neutron-rich isotopes shown in Fig.~\ref{f3}, one observes that the
fragmentation of the PDR is much weaker in nuclei around shell
closure. This confirms our general discussion about fragmentation
above since in the nuclei close to the magic numbers the excitation
energies of the low-lying 2$^+$ states are noticeably higher and the
transition probabilities are reduced compared to the situation in
open-shell nuclei. This means that the respective matrix elements of
the phonon vertices have also smaller values, leading to a weaker
nucleon-vibration coupling.

We further observe that the widths of the GDR are also reduced in
magic nuclei as compared to nuclei with opened shells. While typical
width values are 5-5.5 MeV~\cite{LRT.08} in agreement with data and
R(Q)TBA calculations, they are reduced to 3.4 and 4.0 MeV in the
R(Q)TBA calculations for the doubly magic $^{100}$Sn~\cite{LRT.08}
and $^{132}$Sn isotopes. This reduction has, obviously, a similar
origin as the reduction of the PDR fragmentation discussed above,
confirming a strong resemblance between the GDR and PDR features.

\begin{figure}[ptb]
\begin{center}
\includegraphics*[scale=1.0]{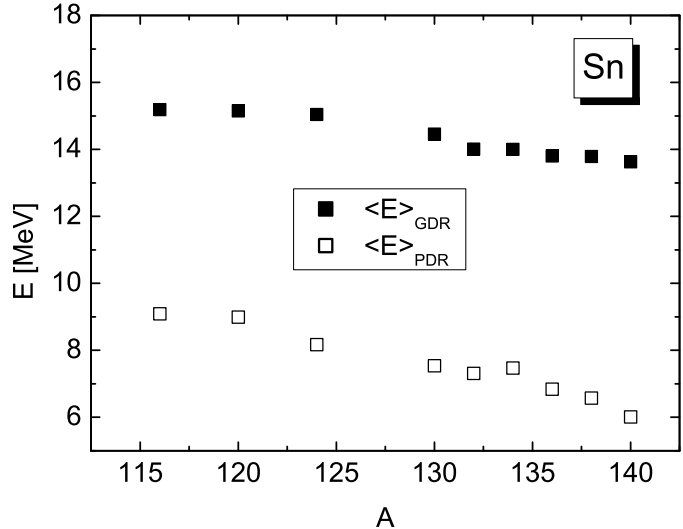}
\end{center}
\caption{The mean energies for the GDR (filled symbols) and the PDR (opened symbols) in tin isotopes
calculated within the RQTBA.}
\label{f4}%
\end{figure}

\begin{figure}[ptb]
\begin{center}
\includegraphics*[scale=1.2]{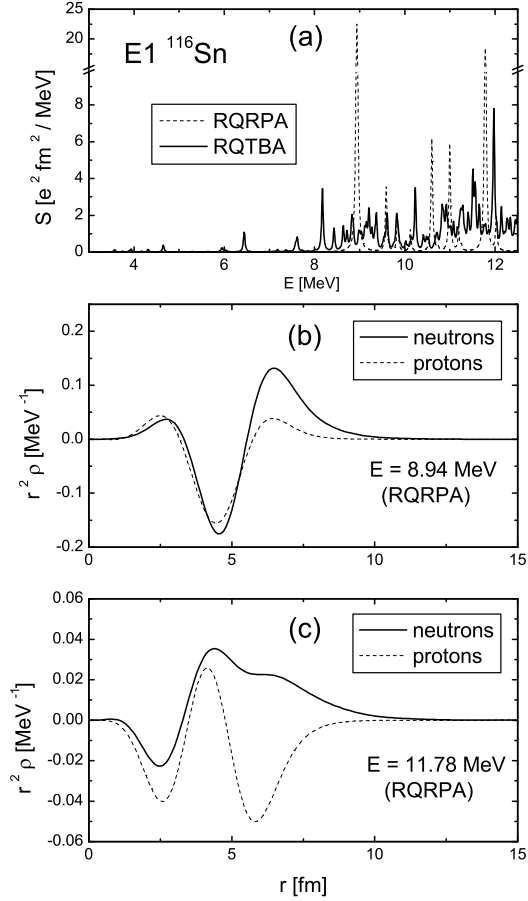}
\end{center}
\caption{The low-lying part of the dipole spectrum in $^{116}$Sn, calculated in
RQRPA (dashed curve) and RQTBA (solid curve) models (upper panel).
RQRPA transitional densities for the most intensive peaks at 8.94 MeV (middle) and 11.78 MeV
(bottom). The dashed curves represent the proton, and the solid curves - the neutron transitional densities.
See text for the detailed explanation.}
\label{f5}%
\end{figure}

\begin{figure}[ptb]
\begin{center}
\includegraphics*[scale=1.2]{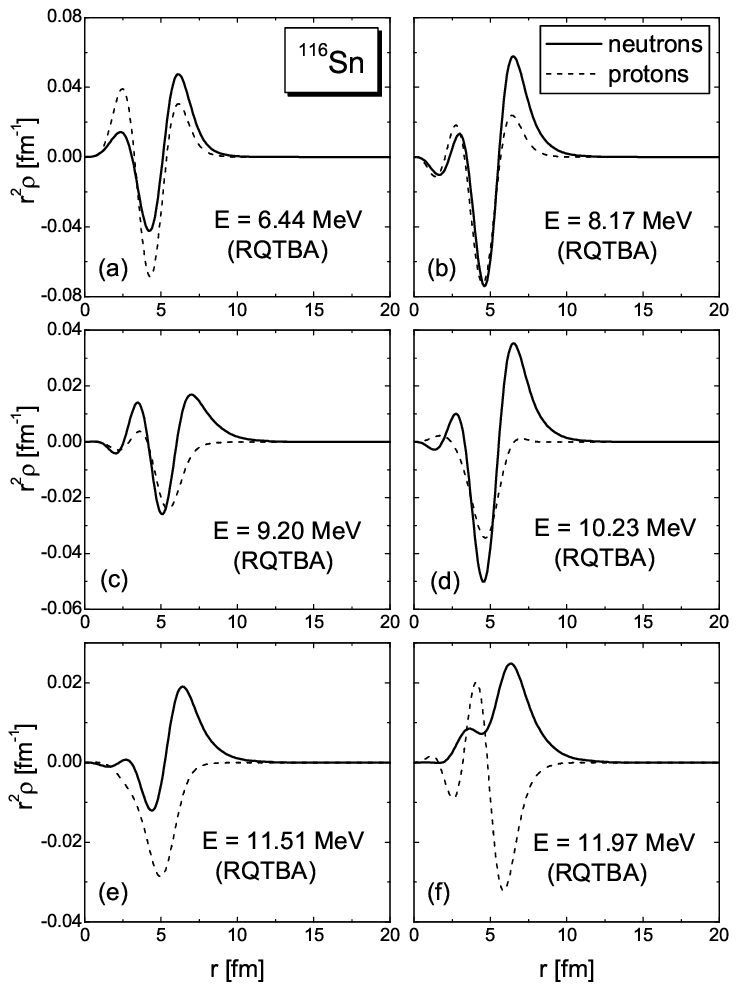}
\end{center}
\caption{The RQTBA transition densities for the low-lying dipole states in $^{116}$Sn.
The dashed curves represent the proton, and the solid curves - the neutron transition densities.}
\label{f6}%
\end{figure}

\begin{figure}[ptb]
\begin{center}
\includegraphics*[scale=1.2]{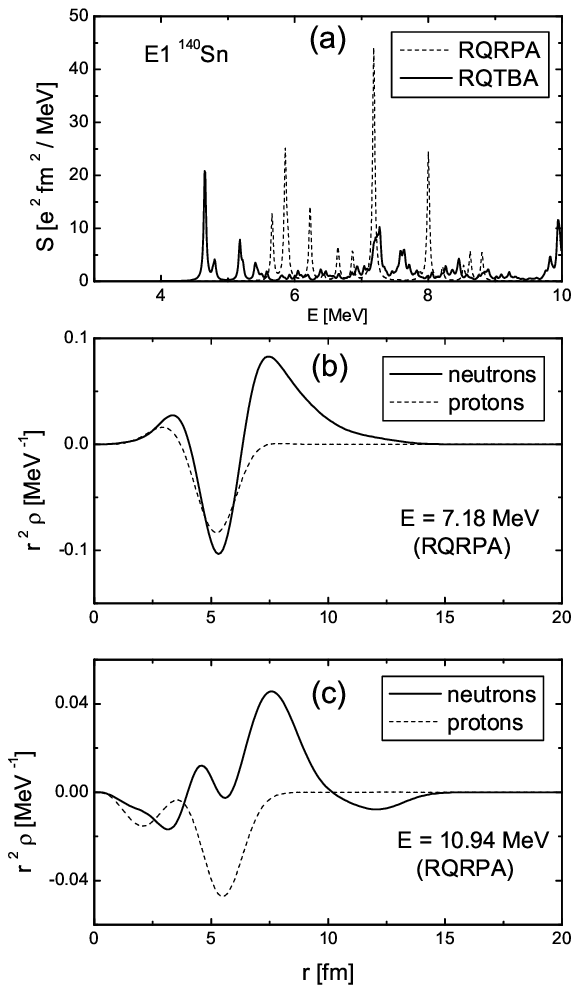}
\end{center}
\caption{Same as in Fig. \ref{f5}, but for $^{140}$Sn.}
\label{f7}%
\end{figure}

\begin{figure}[ptb]
\begin{center}
\includegraphics*[scale=1.2]{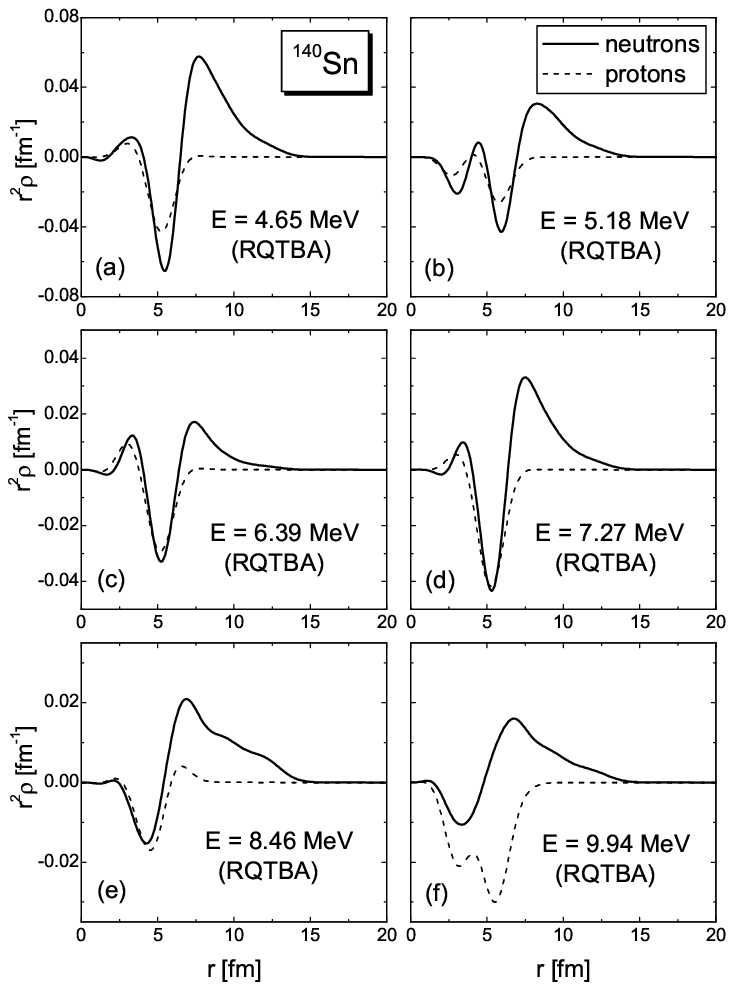}
\end{center}
\caption{Same as in Fig. \ref{f6}, but for $^{140}$Sn.}
\label{f8}%
\end{figure}

The smooth decrease of the GDR mean energy with the nuclear mass
number A, predicted by the well known systematic rule $\langle
E\rangle \simeq 80 A^{-1/3}$, is well reproduced in our RQTBA
calculations. Our results for the GDR and the PDR mean energies are
also in a good agreement with the results of the RQRPA studies of
Ref.~\cite{PNVR.05} for tin and nickel isotopes obtained with the
density-dependent meson-exchange forces (DD-ME1). Visually, the whole
dipole spectrum including both the GDR and the PDR regions shifts
towards lower energies with increasing mass number. However by closer
inspection, we observe that by increasing only the neutron number
and, thus the isospin asymmetry,  the fragmented pygmy resonance
moves down quicker than the GDR. This is illustrated in
Fig.~\ref{f4}. Whereas in $^{116}$Sn and in the isotopes with even
smaller neutron excess the PDR is barely visible in the cross
sections, for more neutron-rich nuclei, starting from $^{130}$Sn it
appears as a pronounced structure which is well separated from the
GDR, in both the RQRPA and RQTBA. This can be understood in terms of
the well known Brown-Bolsterli model \cite{BB.59} where an increase
of collectivity and also of the strength of the lowest solution is
directly connected to its shift towards lower energies. This
qualitative observation about the separation of the PDR from the GDR
in neutron-rich nuclei will be confirmed by the analysis of
transition densities in the energy region around 10 MeV which we
present below (see Figs.~\ref{f5}, \ref{f6}, \ref{f7}, \ref{f8}). In
Fig.~\ref{f3} one can see how the GDR and PDR strengths develop with
a further increase of the neutron excess. The PDR continues its
separation moving to lower energies, and its fragmentation also
becomes stronger as well as the fragmentation of the GDR. The latter
is related to the considerable lowering and increasing collectivity
of the first 2$^+$ states. For example, the RQRPA predicts the first
2$^+$ state at 0.6 MeV in $^{140}$Sn indicating a trend to a static
quadrupole deformation in this nucleus. Indeed, the systematic
deformed relativistic Hartree-BCS calculations with the NL3 force of
Ref.~\cite{LRR.99} indicate a phase-transition from spherical to
axially deformed shapes in this region.

Before we study the structure of the PDR in neutron-rich nuclei in
some details, we would like to make few comments about their
astrophysical importance. As has been pointed out by Goriely and
co-workers~\cite{GKS.04}, an enhancement of the dipole strength at
energies just above the neutron thresholds can increase the neutron
capture cross sections relevant for simulations of the astrophysical
r-process. From the tin isotopes studied here, the double-magic
$^{132}$Sn and the even more neutron-rich nuclei are close to the
r-process path. As can be seen in Fig.~\ref{f2}, our calculation
predicts a sizable contribution of the PDR in $^{132}$Sn just above
the neutron threshold which is, however, relatively high (about 7
MeV) in this double-magic nucleus. The presence of such PDR strength
increases the neutron capture cross section on $^{131}$Sn compared to
the usual treatment which models the low-lying dipole strength by a
Lorentzian fit to the GDR strength~\cite{Cowan.91}. As explained
above the location of the PDR is pushed down in energy with
increasing neutron excess. However, the neutron threshold energy
decreases even faster in the tin isotopes beyond $^{132}$Sn. As a
consequence the presence of the low-lying PDR strength, predicted by
our RQTBA calculations, is not expected to increase the corresponding
neutron capture cross sections as it is located noticeably above the
respective neutron thresholds. To the contrary, the lack of dipole
strength just above the neutron thresholds, as predicted by our RQTBA
calculations, might result in lower capture cross sections than
obtained from the parameterized Lorentzian strength distributions. An
answer to this question will be given by statistical model
calculations which will compare cross sections obtained from our
microscopically derived dipole strengths with those using the
standard Lorentzian approximation. These studies are in progress.

The rather strong first dipole states found in our RQTBA calculations
for the nuclei $^{136,138,140}$Sn can be interpreted as the
prototypes of the 2$^+\otimes$3$^-$ states which are usually observed
in dipole spectra and always appear, for instance, in QPM
calculations (see, for instance, Ref.~\cite{GBB.98}). Since in the
present version of the RQTBA we confine ourselves to $2q\otimes
phonon$ configurations, the position of each peak, if it would be
unperturbed by the static residual interaction, is determined by the
sum of the two quasiparticle energies and the phonon energy, whereas
in the QPM it is determined by the sum of two phonon energies which
is, as a rule, considerably lower in energy and has correspondingly
often an enhanced transition probability. A similar effect should
take place in the two-phonon version of our model where instead of
two unperturbed quasiparticles their bound state, a phonon, is used
as an elementary excitation (see~\cite{Tse.07} for the details). In
the present case the $2q\otimes phonon$ configurations in nuclei with
medium neutron excess, e.g. configurations consisting of two
quasiparticles with total angular momentum and parity 3$^-$ and a
2$^+$ phonon, have higher energies and reduced probabilities, so that
they are not noticeable in the lowest part of the spectrum. In more
neutron-rich, nuclei, however, the 2$^+$ states are pushed down to
such low energies and relatively strong transition probabilities that
their matrix elements are large enough to form a low-lying $2q\otimes
phonon$ configuration with the sizable strength. We note that the
same explanation is applicable to the case of neutron-rich nickel
isotopes to be discussed below (see Figs.~\ref{f8},~\ref{f9}).

For a detailed inspection of the structure of prominent peaks in the
RQRPA and RQTBA dipole strength distributions we have calculated
their respective transition densities. For comparison we focus on two
nuclei: the stable isotope $^{116}$Sn with a rather small neutron
excess (Figs.~\ref{f5},~\ref{f6}) and $^{140}$Sn with a large isospin
asymmetry (Figs.~\ref{f7},~\ref{f8}). In the top panels of figures
\ref{f5} and \ref{f7} we display the low-lying dipole strength
distribution, calculated within the RQRPA and RQTBA. The middle and
bottom panels of these figures exhibit the neutron (solid curves) and
the proton (dashed curves) RQRPA transition densities for the most
pronounced peak of the PDR region (middle) and the lowest peak of the
GDR region (bottom) as obtained within the RQRPA approach. The choice
of the latter peak indicates approximately the energy region where
the PDR can be separated from the low-energy tail of the GDR.

Fig.~\ref{f5} shows that in the RQRPA model nearly the entire pygmy
strength in $^{116}$Sn is concentrated in a single very strong peak
localized at 8.94 MeV. The neutron component obviously dominates the
transition density at the surface, however, with a noticeable proton
admixture (middle panel of Fig.~\ref{f5}). In contrast, the
corresponding panel of Fig.~\ref{f7} implies that the proton
component is fully suppressed at the nuclear surface in the pygmy
mode of $^{140}$Sn, corresponding to the state at 7.18 MeV. The
bottom panels of Figs.~\ref{f5} and~\ref{f7} show that the structure
of the RQRPA peaks changes drastically when the excitation energy
increases and the lower part of the GDR region is reached. Already
the peaks at 11.78 MeV in $^{116}$Sn and at 10.94 MeV in $^{140}$Sn
exhibit a completely different relative behavior of the neutron and
proton components. The two components are still in phase in the
nuclear interior for $^{116}$Sn, while they are not for $^{140}$Sn.
For $^{116}$Sn neutrons and protons are out of phase at the surface
with protons even dominating, whereas in $^{140}$Sn protons do not
contribute to the surface motion. Thus the bottom panels indicate
that the states at these moderate excitation energies are no longer
of pygmy type, but do also not yet exhibit the typical GDR structure
with protons and neutrons oscillating against each other.

In Figs.~\ref{f6} and~\ref{f8} we display the dipole transition
densities for selected states in $^{116}$Sn and $^{140}$Sn, as
calculated within the RQTBA. These states arise from the coupling of
the RQRPA states to 2$^+$, 3$^-$, 4$^+$, 5$^-$, 6$^+$ phonons with
energies below 10 MeV. As it has been shown in
Refs.~\cite{LRT.07,LRV.07} for the case without pairing correlations,
the RTBA transition densities obey more general normalization
condition than the pure RPA states. This normalization condition is
generalized straightforwardly to systems with pairing correlations of
the superfluid type. Formally, this condition in the RQTBA has the
same form as in the RTBA, but the respective summation is performed
over the doubled quasiparticle space. In the notations of Ref.
\cite{LRT.08}, the deviation of the quantity
$\sum\limits_{(k_1k_2)\eta}\eta|{\cal R}^{\eta}_{\mu(k_1k_2)}|^2 $
(where ${\cal R}^{\eta}_{\mu(k_1k_2)}$ are the matrix elements of the
transition densities of the state $\mu$) from unity indicates
quantitatively the effect of the coupling to phonons of the state
$\mu$. Notice, that the amplitudes of oscillations are reduced in the
RQTBA as compared to the RQRPA as reflected in Figs. \ref{f6} and
\ref{f8}. These figures give a detailed look into the underlying
structure of the RQTBA dipole spectrum. In particular, we observe
that in the case of strong configuration mixing, like in the RQTBA,
both nuclei develop a transitional region with states which have
neither pure isoscalar nor isovector underlying structures. This is
in strong contrast to the RQRPA which does not show such a smooth
transition range. For example, except for two relatively weak
excitations around 9 MeV, the RQRPA spectrum for $^{140}$Sn does not
contain states in the energy range between 8 and 11 MeV. Thus we
conclude that the RQRPA predicts a clear separation between the PDR
and GDR regions. This separation is, however, overcome in the RQTBA
due to strong mode mixing giving rize to the formation of a
transitional region between PDR and GDR containing many weak states
with mixed structure.

\begin{table}[ptb]
\caption{Distribution of two-quasiparticle configurations contributing to the low-lying dipole
states in $^{116}$Sn calculated within the RQRPA and the RQTBA.}
\label{tab116}
\begin{center}
\vspace{6mm}
\tabcolsep=0.2em \renewcommand{\arraystretch}{1.2}%
\begin{tabular}
[c]{c|c|c|c}
\hline\hline
  RQRPA, 8.94 MeV & RQTBA, 6.44 MeV & RQTBA, 9.20 MeV & RQTBA, 10.23 MeV\\
\hline
 30.8\% (2d$_{5/2}$ $\to$ 2f$_{7/2}$)$\nu$   & 19.6\% (2p$_{3/2}$ $\to$ 2d$_{5/2}$)$\pi$  & 8.6\% (2d$_{3/2}$ $\to$ 3p$_{1/2}$)$\nu$ & 6.1\% (2d$_{5/2}$ $\to$ 2f$_{7/2}$)$\nu$ \\
 16.9\% (1g$_{9/2}$ $\to$ 1h$_{11/2}$)$\nu$  &  9.7\% (1g$_{9/2}$ $\to$ 1h$_{11/2}$)$\pi$ & 1.0\% (2p$_{1/2}$ $\to$ 3s$_{1/2}$)$\pi$ & 3.2\% (1g$_{7/2}$ $\to$ 2f$_{5/2}$)$\nu$ \\
 14.2\% (1g$_{9/2}$ $\to$ 1h$_{11/2}$)$\pi$  &  0.5\% (1g$_{7/2}$ $\to$ 2f$_{5/2}$)$\nu$  & 1.0\% (2d$_{3/2}$ $\to$ 3p$_{3/2}$)$\nu$ & 1.0\% (2d$_{3/2}$ $\to$ 2f$_{5/2}$)$\nu$ \\
  8.7\% (2p$_{3/2}$ $\to$ 2d$_{5/2}$)$\pi$   &  0.5\% (2d$_{5/2}$ $\to$ 2f$_{7/2}$)$\nu$  & 0.8\% (2d$_{5/2}$ $\to$ 3p$_{3/2}$)$\nu$ & 0.8\% (1g$_{9/2}$ $\to$ 1h$_{11/2}$)$\nu$ \\
  6.1\% (1g$_{7/2}$ $\to$ 2f$_{5/2}$)$\nu$   &  0.4\% (1f$_{5/2}$ $\to$ 2d$_{3/2}$)$\pi$  & 0.8\% (2p$_{3/2}$ $\to$ 3s$_{1/2}$)$\pi$ & 0.8\% (2p$_{1/2}$ $\to$ 3s$_{1/2}$)$\nu$ \\
  3.4\% (2d$_{5/2}$ $\to$ 3p$_{3/2}$)$\nu$   &  0.2\% (1f$_{5/2}$ $\to$ 2d$_{5/2}$)$\pi$  & 0.4\% (2d$_{5/2}$ $\to$ 2f$_{7/2}$)$\nu$ & 0.6\% (1g$_{7/2}$ $\to$ 2f$_{7/2}$)$\nu$ \\
  3.1\% (1g$_{7/2}$ $\to$ 2f$_{7/2}$)$\nu$   &  0.1\% (1g$_{9/2}$ $\to$ 1h$_{11/2}$)$\nu$ & 0.4\% (1g$_{7/2}$ $\to$ 1h$_{9/2}$)$\nu$ & 0.5\% (2p$_{3/2}$ $\to$ 3s$_{1/2}$)$\nu$ \\
  2.5\% (2p$_{1/2}$ $\to$ 2d$_{3/2}$)$\pi$   &  0.1\% (2p$_{1/2}$ $\to$ 2d$_{3/2}$)$\pi$  & 0.3\% (2p$_{1/2}$ $\to$ 2d$_{3/2}$)$\pi$ & 0.5\% (1h$_{11/2}$ $\to$ 1i$_{13/2}$)$\nu$ \\
  1.7\% (1f$_{5/2}$ $\to$ 1g$_{7/2}$)$\pi$   &                                            & 0.2\% (3s$_{1/2}$ $\to$ 3p$_{3/2}$)$\nu$ & 0.5\% (2p$_{3/2}$ $\to$ 2d$_{3/2}$)$\nu$ \\
  1.4\% (1f$_{5/2}$ $\to$ 2d$_{3/2}$)$\pi$   &                                            & 0.2\% (1g$_{9/2}$ $\to$ 2f$_{7/2}$)$\pi$ & 0.3\% (1g$_{7/2}$ $\to$ 1h$_{9/2}$)$\nu$ \\
  1.0\% (3s$_{1/2}$ $\to$ 3p$_{3/2}$)$\nu$   &                                            & 0.2\% (3s$_{1/2}$ $\to$ 3p$_{1/2}$)$\nu$ & 0.3\% (1g$_{9/2}$ $\to$ 1h$_{11/2}$)$\pi$ \\
                                             &                                            & 0.1\% (1f$_{5/2}$ $\to$ 1g$_{7/2}$)$\pi$ & 0.2\% (1f$_{5/2}$ $\to$ 2d$_{3/2}$)$\pi$ \\
                                             &                                            & 0.1\% (2d$_{5/2}$ $\to$ 2f$_{5/2}$)$\nu$ & 0.1\% (2p$_{1/2}$ $\to$ 3s$_{1/2}$)$\pi$ \\
                                             &                                            &                                          & 0.1\% (1f$_{5/2}$ $\to$ 1g$_{7/2}$)$\pi$ \\
\hline\hline
\end{tabular}
\end{center}
\end{table}

Table \ref{tab116} presents distributions of two-quasiparticle
configurations which contribute to the selected low-lying dipole
states in $^{116}$Sn: in the first column we display the
configurations forming the RQRPA state at 8.94 MeV which
accumulates almost the entire strength with isoscalar underlying
structure within the RQRPA. From Table \ref{tab116} one can see
that this is a collective state since  many $2q$ configurations
contribute with comparable strengths. From this distribution one
can also conclude that in this nucleus, although neutron
configurations dominate the pygmy mode, admixtures of proton
configurations are still sizable, since the neutron excess is not
yet large. As it has been discussed above, in the RQTBA this state
is strongly coupled to phonons and fragments within the energy
region between 4 and 10 MeV into many states with isoscalar
underlying structure. The second, third and fourth columns of
Table \ref{tab116} represent the $2q$ contributions to the states
at 6.44 MeV, 9.20 MeV and 10.23 MeV, respectively. Note, that the
total percentage of the $2q$ configurations is very much reduced
with respect to the 100 \% normalization which one has in the
RQRPA. This reduction from unity is a measure of the coupling
strength to phonons for a certain state.  We find that for the
considered states in this region deviations are about 50-70 \%
that indicates very strong coupling.

\section{RQTBA studies for neutron-rich nickel isotopes}
\label{ni}

In this section we present the results of our studies of the
dipole strength in neutron-rich nickel isotopes. One of our
motivations for choosing these isotopes is the recent experimental
work \cite{Bra.07} where the low-lying dipole strength in
$^{68}$Ni has been studied by Coulomb dissociation, observing a
pronounced and broad bump in the excitation spectrum with a
centroid between 10 and 11 MeV. Our second motivation stems from
nuclear astrophysics as the double-magic nucleus $^{78}$Ni is
considered the essential waiting point at the $N=50$ shell closure
for certain r-process nucleosynthesis simulations \cite{Cowan.91}.

\begin{figure}[ptb]
\begin{center}
\includegraphics*[scale=1.2]{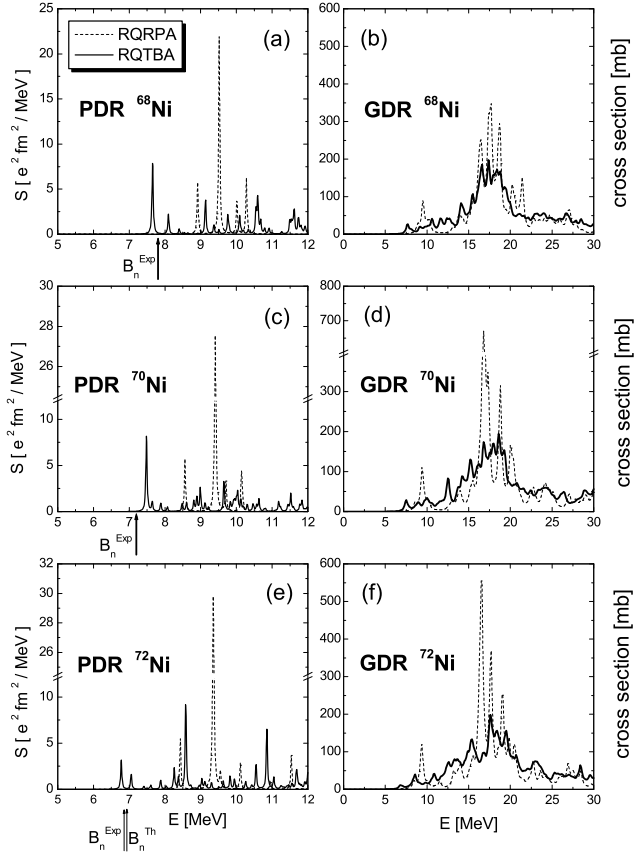}
\end{center}
\caption{Same as in Fig. \ref{f2}, but for $^{68,70,72}$Ni isotopes.}
\label{f9}%
\end{figure}

\begin{figure}[ptb]
\begin{center}
\includegraphics*[scale=1.2]{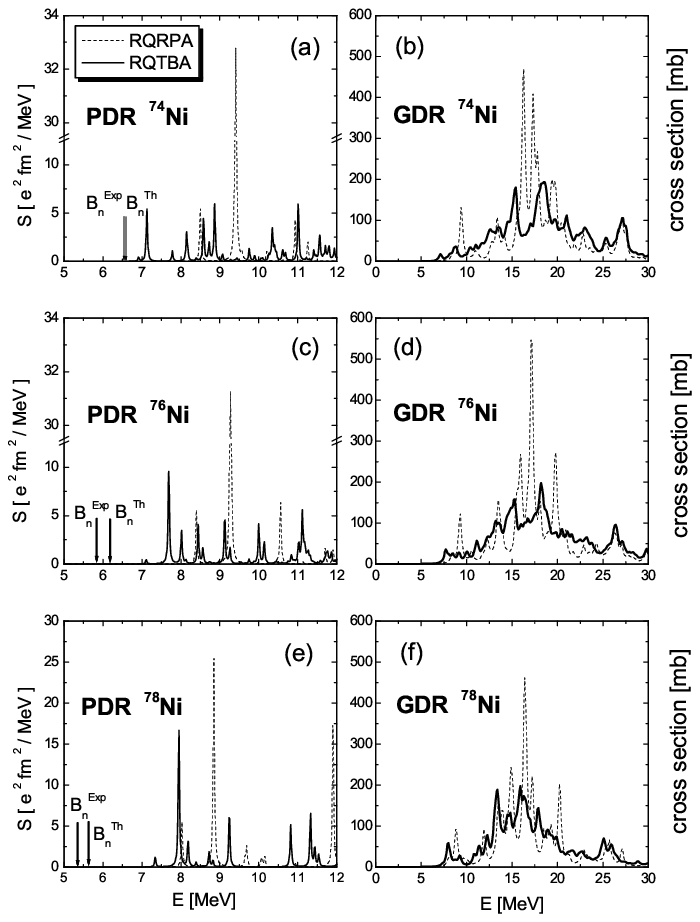}
\end{center}
\caption{Same as in Fig. \ref{f2}, but for $^{74,76,78}$Ni isotopes.}
\label{f10}%
\end{figure}

Analogously to the presentations of the tin isotopes in Figs.
\ref{f2} and \ref{f3}, we display the RQRPA and RQTBA dipole
spectra for the even nickel isotopes $^{68}$Ni -- $^{78}$Ni in
Figs. \ref{f9} and \ref{f10}. The overall pattern of the dipole
spectra (right panels), including the fragmentation of the GDR and
of the low-energy part of the spectra, is quite similar for
$^{68}$Ni and $^{70}$Ni, which differ only by two neutrons in the
$1g_{9/2}$ orbit within the independent particle model. The
similarity of the fragmentation effects is caused by the fact that
the  first 2$^+$ states, which are mainly responsible for the
fragmentation, have comparable energies and transition
probabilities in the RQRPA for both nuclei. In $^{72,74,76}$Ni the
PDR and GDR are stronger fragmented due to lowering of the
energies and an accompanying increase of the transition
probabilities of the first 2$^+$ states. As expected, in the
double-magic nucleus $^{78}$Ni the fragmentation is reduced again.

Let us discuss our results for $^{68}$Ni in more details. For the
low-lying part of the dipole spectrum (Fig.~\ref{f9}(a)), the RQRPA
calculation (dashed curve) predicts one rather pronounced peak at
9.52 MeV, accompanied by three relatively weak states. The transition
density of the main peak has the isoscalar underlying structure
typical for the pygmy mode (see Fig.~\ref{f11} below). The enlarged
picture of the low-energy part of the spectrum, calculated with the
smaller smearing parameter (upper panel of Fig. \ref{f11}) indicates
that the next RQRPA structures appear only at around 14 MeV. The
transition densities of these states do not have an isoscalar
underlying structure in the nuclear interior, but behave already as
the low-energy tail of the GDR. Thus, in this nucleus, as well as for
the heavier nickel isotopes (Figs.~\ref{f9} and \ref{f10}), we
observe the pronounced separation in the RQRPA dipole strength
distribution between the PDR and the GDR regions, as already familiar
from the neutron-rich tin isotopes.

\begin{figure}[ptb]
\begin{center}
\includegraphics*[scale=1.2]{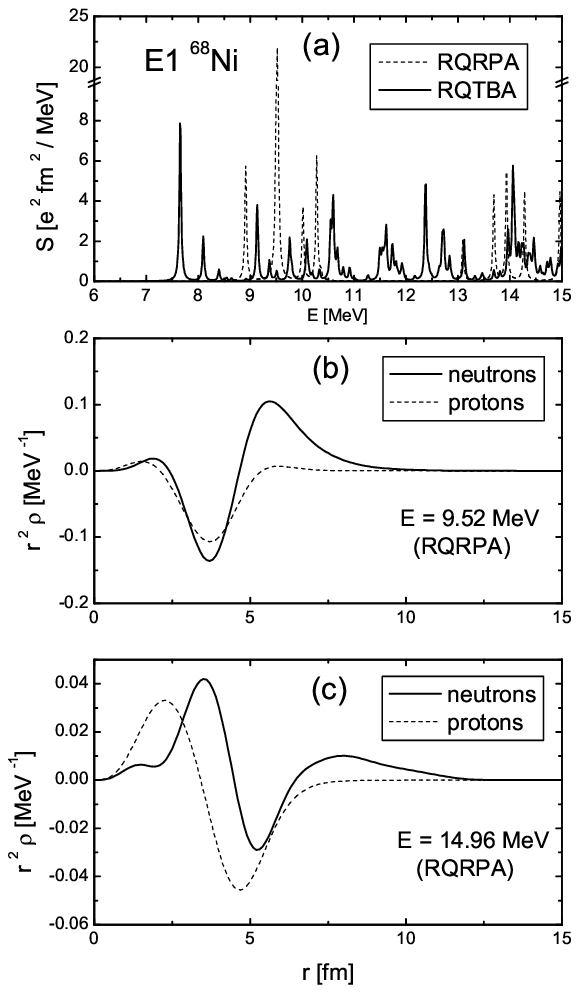}
\end{center}
\caption{Same as in Fig. \ref{f5}, but for the $^{68}$Ni.}
\label{f11}%
\end{figure}

\begin{figure}[ptb]
\begin{center}
\includegraphics*[scale=1.2]{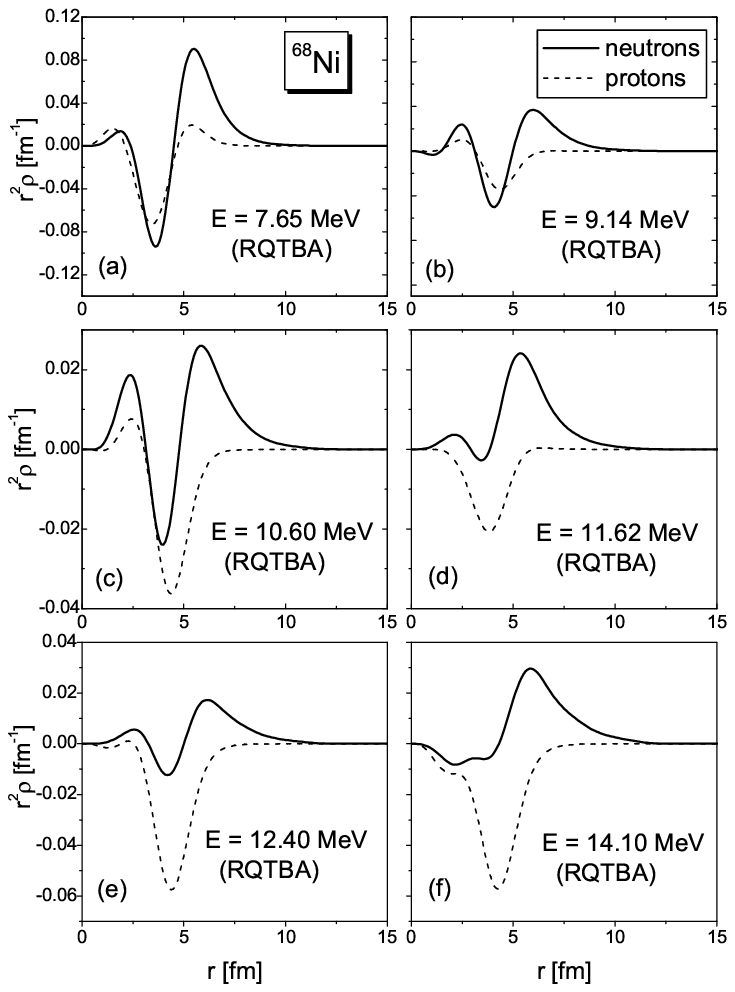}
\end{center}
\caption{Same as in Fig. \ref{f6}, but for the $^{68}$Ni.}
\label{f12}%
\end{figure}

When coupled to phonons the pygmy mode, which is localized in the
RQRPA dipole spectrum for $^{68}$Ni between 9 and 10.5 MeV, spreads
over the broad energy region from 7.5 to about 12 MeV. The situation
is very similar in the other investigated nickel isotopes. The
corresponding RQTBA strength distributions in Figs.~\ref{f9},
\ref{f10} show many states with comparable strength in the energy
region below 12 MeV. One of the lowest sharp peaks in $^{68}$Ni --
$^{76}$Ni is supposed to have a $3^-\otimes2^+$ $2q\otimes phonon$
nature. Hence the energy of such a state is expected to be pushed
down in the two-phonon version of the RQTBA. Among the single peaks
at low energies in $^{68}$Ni one can see a concentration of strength
centered at 10.60 MeV which might correspond to the experimentally
observed peak in Ref.~\cite{Bra.07}. In Fig. \ref{f12} the transition
densities of selected RQTBA states in the low-lying region are
displayed. One can see that, in the case of $^{68}$Ni only the few
lowest states exhibit a pure isoscalar underlying structure. In the
energy region between 10.5 and 13.5 MeV, for which the RQRPA does not
produce any dipole strength, the RQTBA predicts a large amount of
states with transitional behavior: their density distributions have
neither pure isoscalar nor isovector underlying structure, however,
the isoscalar character dominates below 12 MeV.

Astrophysically, $^{78}$Ni plays an important role for the matter
flow beyond the $N=50$ shell closure in certain r-rocess scenarios.
This has been demonstrated in Ref.~\cite{Hosmer.05} reporting the
first measurement of the $^{78}$Ni half life which turned out to be
noticeably shorter than predicted by the global models and had a
decisive influence of the local r-process abundance distribution
around the $N=50$ nuclides. Our focus here is the question whether
our microscopic approach predicts an enhanced dipole strength at
energies just above the neutron threshold which might affect the
neutron capture cross section to $^{78}$Ni. As can be seen in
Fig.~\ref{f10}, our RQTBA calculation indeed predicts the existence
of a strong pygmy strength in $^{78}$Ni, which, however, resides more
than 2 MeV above the neutron threshold and hence is expected to have
only a mild effect on the capture cross sections. Nevertheless,
statistical model calculations employing our RQTBA dipole strength
distributions are in progress for all nickel isotopes.

As a more general remark we note that our RQTBA results should be
reliable enough to check whether the Lorentzian approximation to
the dipole strength function as implied from data for stable
nuclei and often employed to derive the relevant neutron capture
cross sections for r-process nuclei is also valid for nuclei with
large neutron excess. Such studies are also in progress.

\section{Systematic of the pygmy dipole strength}

\begin{figure}[ptb]
\begin{center}
\includegraphics*[scale=1.0]{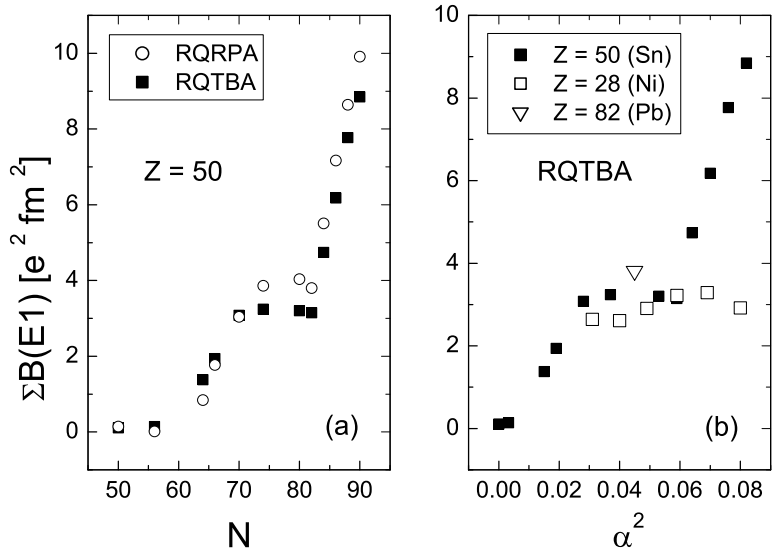}
\end{center}
\caption{Integrated dipole strength with isoscalar underlying
structure below GDR (pygmy strength) calculated within the RQTBA.
Left panel (a): pygmy strength in tin isotopes, calculated within
the RQRPA (open circles) and the RQTBA (solid squares), versus the
neutron number. Right panel (b): pygmy strength in nuclei versus
the squared asymmetry parameter $\alpha = (N-Z)/A$ in Sn isotopes
(solid squares), Ni isotopes (open squares), and $^{208}$Pb
(triangle).}
\label{f13}%
\end{figure}

This section is devoted to a study of the systematics of the total
PDR strength as a function of neutron excess. Obviously, the
crucial point of such an investigation is the determination of the
energy at which the PDR and GDR regions separate; i.e. the energy
at which the underlying structure of the states in the dipole
strength function change from an isoscalar to an isovector
character. We have determined this energy by a careful inspection
of the transition densities of the individual states and their
underlying structure. Having determined in this way the upper
energy limit, we have calculated the total PDR strength by
integration of the strength function obtained within the RQRPA and
the RQTBA with a smearing parameter of 20 keV. The left panel of
Fig. \ref{f13} displays the integrated RQRPA and RQTBA pygmy
strength $\sum B(E1)\uparrow$ for the tin isotopes versus their
neutron number. In the right panel one can see the integrated
pygmy strength calculated within the RQTBA for tin and nickel
isotopes as well as for $^{208}$Pb as functions of the squared
asymmetry parameter $\alpha = (N-Z)/A$.

For the tin isotopes the total pygmy strength generally increases
with neutron number, both in the RQRPA and RQTBA. One can see from
Fig.~\ref{f13}(a) that the difference between the RQRPA and the RQTBA
results for the integrated pygmy strength is not sizable. This
confirms our general suggestion that the phonon coupling causes
fragmentation and redistribution of the pygmy mode without breaking
its integrity, as it happens with the GDR. Some difference between
the RQRPA and the RQTBA total pygmy strength may be explained by
uncertainties in determining of the upper integration energy limit.
Being well determined in the RQRPA, this quantity is only approximate
in the RQTBA case because for states with the transitional underlying
structure we rely only on our visual impression deciding if a certain
state has more isoscalar or more isovector underlying structure.
Nevertheless, in the present study we have determined the value 10
MeV as the upper integration limit for the pygmy strength in tin
isotopes, 12 MeV -- for nickel isotopes and 8 MeV -- for $^{208}$Pb.

However, the behavior of the integrated pygmy strength shown in
Fig.~\ref{f13} is not purely linear: our RQTBA calculations predict a
nearly constant pygmy strength for the tin isotopes $^{120}$Sn --
$^{132}$Sn, while the RQRPA pygmy strength shows also a deviation
from the linear behavior within this isotopic range. Note, that these
isotopes have in common that the the last (partially) occupied
neutron subshell is the $1h_{11/2}$ orbit, which is closed at N=82.
In the right panel, where we present the RQTBA pygmy strength versus
the asymmetry parameter, one can see that the total pygmy strength in
the neutron-rich nickel isotopes $^{68}$Ni-$^{78}$Ni demonstrates a
very similar behavior: it remains almost constant for these isotopes.
Interestingly, for these nuclei the neutron $1g_{9/2}$ subshell is
also being filled up closing the N=50 shell. In addition, the
$1g_{9/2}$ orbit and the $1h_{11/2}$ orbit are intruder orbits in the
Ni isotopes and Sn isotopes, respectively.

Another interesting observation is that the absolute values of the
pygmy strength in tin and nickel isotopes with the same value of
the asymmetry parameter are roughly close to each other and even
to the value deduced for $^{208}$Pb from the calculations of Ref.
\cite{LRT.07}. We observe quite a big difference between the total
pygmy strength in Sn and Ni chains only at $\alpha^2 > 0.06$ that
can be explained by approaching the shell closure N = 50 in the
nickel chain. Certainly, to make a general statement about the
dependence of the pygmy strength on the asymmetry parameter, one
should perform a more wide investigation of different isotopic
chains, but our observation agrees, in general, with the
observation of Ref. \cite{KPA.07} based on the systematic of the
available experimental data.

Notice, that in the present studies we use the NL3 interaction which
is known to overestimate systematically the neutron skin thickness
\cite{NVFR.02}. The reason is that in the NL3 parameter set the meson
coupling in the isovector channel has no density dependence and is
parameterized by only one coupling constant. As a result, the total
pygmy strength in nuclei with large neutron excess can be
respectively overestimated as compared, for instance, to the values
obtained in the RQRPA calculations \cite{PNVR.05} with DD-ME2
interaction presented in Ref. \cite{OEN.07}.

 From our systematic studies we conclude that in the models
based on the CEDFT the total pygmy strength can deviate from the
linear dependence on the neutron number or on the asymmetry
parameter. In Refs. \cite{HBK.04}, \cite{TTK.07} it has been found
within a non-relativistic model with Landau-Migdal interaction,
that the coupling to collective vibrations produces a noticeable
deviation of the pygmy dipole strength from the linear dependence
on the neutron number in calcium isotopes, in agreement with the
experimental data of Ref. \cite{HBK.04}. From our present
investigation for the tin chain we conclude that this deviation
can take place around the shell closures with intruder orbits both
for the RQRPA and the RQTBA systematics.

\section{Summary}
\label{summary}

The relativistic quasiparticle time blocking approximation
developed recently \cite{LRT.08} is applied to a systematic study
of dipole spectra in even-even neutron-rich tin and nickel nuclei
with a special focus on the low-lying part of the dipole strength.

First of all, a very reasonable agreement of the RQTBA results with
recent measurements for $^{130}$Sn, $^{132}$Sn~\cite{Adr.05}, which
determined the dipole strength above the neutron thresholds, is
found. The RQTBA is also applied to calculations for nuclei with
large isospin asymmetry paying special attention to the fragmentation
of the GDR and the PDR modes due to coupling to phonons. It has been
found that the fragmentation is considerably reduced in doubly-magic
nuclei and in some cases in neighboring even-even ones. The
connection of this effect with the properties of the lowest $2^+$
states has been discussed.

The transition densities of the low-lying dipole states have been
computed within the RQRPA and the RQTBA. It has been shown that the
pygmy mode with the isoscalar underlying structure spreads in the
broad energy region over many states whose underlying structure
demonstrates a similar type of behavior which helps to identify the
fragments of the pygmy mode and to separate them from the low-energy
GDR tail. With help of this analysis the integral pygmy strengths in
the investigated nuclei have been computed. It has been shown that
the calculated integral pygmy strength demonstrates a nearly linear
behavior versus the asymmetry parameter with some deviations from
this behavior near the shell closures.

As a next step to astrophysical applications, the RQTBA dipole
strength distributions will be used in statistical model
calculations to derive neutron capture cross sections for
r-process simulations.

\bigskip\leftline{\bf ACKNOWLEDGEMENTS}

The authors are thankful to Adam Klimkiewicz for performing the
convolution of our dipole strength functions for $^{130,132}$Sn
with the LAND-FRS detector response and providing us with the
experimental data. Helpful discussions with T. Aumann, K.
Boretzky, I. Borzov, H. Feldmeier are gratefully acknowledged.
This work has been supported in part by the Bundesministerium
f\"{u}r Bildung und Forschung under project 06 MT 246 and by the
DFG cluster of excellence
\textquotedblleft Origin and Structure of the Universe\textquotedblright%
\ (www.universe-cluster.de). V.~T. acknowledges financial support
from the Deutsche Forschungsgemeinschaft under the grant No. 436
RUS 113/806/0-1 and from the Russian Foundation for Basic Research
under the grant No. 05-02-04005-DFG\_a.
\bigskip


\end{document}